\documentclass{jnmp}

\usepackage{amsmath}

\JNMPnumberwithin{equation}{section}

\newtheorem{proposition}{Proposition}

\theoremstyle{definition}
\newtheorem{definition}{Definition}
\newtheorem{example}{Example}

\begin{document}

\renewcommand{\evenhead}{M G{\" u}rses, A Karasu and R Turhan}
\renewcommand{\oddhead}{Time-Dependent Recursion Operators and Symmetries}

\thispagestyle{empty}

\FirstPageHead{9}{2}{2002}{\pageref{gurses-firstpage}--\pageref{gurses-lastpage}}{Article}

\copyrightnote{2002}{M G{\" u}rses, A Karasu and R Turhan}

\Name{Time-Dependent Recursion Operators\\
and Symmetries}
\label{gurses-firstpage}

\Author{M G{\" U}RSES~$^\dag$,  A KARASU~$^{\ddag}$ and R TURHAN~$^{\ddag}$}

\Address{$^\dag$~Department of Mathematics, Faculty of Sciences, \\
~~Bilkent University, 06533 Ankara,  Turkey \\
~~E-mail: gurses@fen.bilkent.edu.tr\\[10pt]
$^\ddag$~Department of Physics, Faculty of Arts and Sciences, \\
~~Middle East Technical University, 06531 Ankara, Turkey  \\
~~E-mail: karasu@metu.edu.tr and ref\mbox{}ik@metu.edu.tr}

\Date{Received November 18, 2001; Revised  December 7, 2001;
Accepted January 1, 2002}

\begin{abstract}
\noindent
The recursion operators and symmetries of nonautonomous, $(1+1)$ dimensional
integrable evolution equations are considered. It has been previously
observed that the symmetries of the integrable evolution equations
obtained through their recursion operators do not satisfy the symmetry
equations. There have been several attempts to resolve this problem.
It is shown that in the case of time-dependent evolution equations
or time-dependent recursion operators associativity is lost.
Due to this fact such recursion operators need modification. A general
formula is given for the missing term of the recursion operators. Apart
from the recursion operators  a method is introduced to calculate the
correct symmetries. For illustrations several examples of scalar
and coupled system of equations are considered.
\end{abstract}

\section{Introduction}
Time-dependent local and nonlocal symmetries for autonomous and
nonautonomous integrable equations have been extensively studied
in literature \cite{san}--\cite{kar2}. In a recent
paper~\cite{san} by Sanders and Wang it was observed that
time-dependent recursion operators associated with some integrable
(1+1) dimensional evolution equations do not always generate the
higher order symmetries. They explained this fact by the violation
of rule $D^{-1}D=1$ where $D=D_{x}$. In order to overcome this
problem, they presented a method for constructing symmetries of a
given integrable evolution equation by a {\it corrected recursion
operator} resulting from the {\it (weak)} standard one. In this
paper we consider the work of~\cite{san}, but from a new point of
view. In fact, we show that an elegant way of understanding this
problem is through the action of $D^{-1}$ on arbitrary functions
depending on dependent and independent variables and the structure
of symmetries of equations. On the other hand we investigate the
behavior of recursion operator under a simple Lie point
transformation which links evolution equations. For instance the
cylindrical Korteweg-de Vries (cKdV) equation is related to the
Korteweg-de Vries (KdV) equation by a simple point transformation
which allows also a direct construction of symmetries and
recursion operators for cKdV from those of KdV. The properties of
this transformation or {\it the principle of cova\-riance} implies
that recursion operators must keep their property of {\it mapping
symmetries to symmetries}. The corresponding Lie point
transformation maps symmetries of the KdV equation to the correct
symmetries of the cKdV equation. On the other hand it maps the
recursion operator to the weak one. This fact suggests that under
this transformation $D^{-1} \rightarrow D^{-1}+h(t)$, where $h(t)$
is a time-dependent function to be determined. If such an
integration constant is missed we loose simply the rule of
associativity. As a simple example, let $R_{0}=D^{-1}$,
${K^{\prime}}_{0}=D^2$, and $\sigma_{0}= a_{1}(t)+a_{2}(t)
x+a_{3}(t) x^2+ a_{4}(t) x^3+ \cdots$. Observe that
$R_{0}\,({K^{\prime}}_{0}
(\sigma_{0}))-(R_{0}\,{K^{\prime}}_{0})\, (\sigma_{0}) =-a_{2} \ne
0$. This integration function can be determined either by using
the definition of the recursion operator or the symmetry equation.
In this work we use both approaches.

Most of the nonlinear integrable evolution equations, in $(1+1)$ dimensions,
admit recursion operators which map symmetries to symmetries. Let $A$ be the
space of symmetries of an evolution equation. We assume all symmetries
$\sigma \in A$ are differentiable. This space contains two types of
functions. Let $A_{1}$ be a subset of $A$ containing all functions which vanish in
the limit when the jet coordinates go to zero and $A_{0}$ be a subset of $A$
the elements of which do not vanish under such a limit. A recursion operator ${\cal R}
$ is an operator which maps $A$ into itself ${\cal R}: A \rightarrow A$.
This may be implied by the eigenvalue equation ${\cal R} \sigma= \lambda
\sigma$, where $\sigma \in A$ and $\lambda$ is the spectral constant. The
recursion operators are in general nonlocal operators and the usual
characterization of such operators ${\cal R}$ of system of evolution
equations
\begin{equation}
q^{i}_{t}=K[x,t,q^{j}], \qquad i,j=1,2,\ldots,N,  \label{a0}
\end{equation}
\noindent where $K$ is a locally defined function of $q^{i}$ and its
$x$-derivatives, is given by the equation~\cite{olv}
\begin{equation}
{\cal R}_{t}=[K^{\prime},{\cal R}],  \label{a1}
\end{equation}
where the operator $K^{\prime}$ is the Frech{\'e}t derivative of $K
$. A function $\sigma$ is called a symmetry of (\ref{a0}) if it satisfies
the linearized equation
\begin{equation}
\sigma_{t}=K^{\prime}\sigma .  \label{a2}
\end{equation}
The relation among the symmetries is given by
\begin{equation}
\sigma_{n+1}={\cal R}\sigma_{n} ,\qquad  n=0,1,2,\ldots,  \label{a3}
\end{equation}
which guarantees the integrability of the equation under study. We note that
in \cite{san}, the operator ${\cal R}$ is called a {\it weak recursion
operator} of (\ref{a0}) if it satisfies (\ref{a1}) using the rule $D^{-1}D=1$
. In calculating symmetries of an equation with $K$ and recursion operator
depending explicitly on $x$ and $t$, we observed that the problem arises
when the coefficient of $D^{-1}$ in the recursion operators contains
functions depending only on $x$ and $t$ (no $q$ and its derivatives). As an
example, where the function $K$ and the recursion operator depend explicitly
on $x$ and $t$ but the symmetries can be calculated as usual, we have
\begin{equation}
u_{t}= u_{3x}+{\frac{1 }{2t}}uu_{x}-{\frac{1 }{2t}}xu_{x},
\end{equation}
where the corresponding recursion operator is given by
\begin{equation}
{\cal R}=tD^{2}+{\frac{1 }{3}}u +{\frac{1 }{6}}u_{x}D^{-1}.
\end{equation}
 Keeping the notion of covariance in mind and having the recursion
operators found from~(\ref{a1}) at hand we have to reconsider the action of
the operator $D^{-1}$ in the case of function space with elements that have
explicit $t$ and $x$ dependencies. In the next section we discuss the
principles of covariance by computing the symmetries and recursion operator
of cKdV from those of KdV using an invertible Lie point transformation.

\section{The link between KdV and cKdV}

It is well known that the KdV and cKdV equations are equivalent since their
solutions are related by a simple Lie point transformation \cite{fuc,blu}.
This transformation allows also a direct transfer of symmetries in an
invariant way. Therefore the invertible point transformation,
\begin{equation}
\tau= -2 t^{-1/2},\qquad \xi=x t^{-1/2}, \qquad v=tu+{\frac{1 }{2}}x,
\label{a7}
\end{equation}
takes the cKdV equation
\[
u_{t}=u_{3x}+uu_{x}-{\frac{u }{2t}}
\]
to the KdV equation
\[
v_{\tau} =v_{\xi\xi\xi}+vv_{\xi}.
\]
Thus we can derive the symmetries of cKdV from those of KdV using the
transformation above. The relation between the symmetries of KdV and cKdV, from
above transformation, is given as
\begin{equation}
\delta v=t \delta u.  \label{a8}
\end{equation}
The first four symmetries of KdV equation, generated by
\begin{equation}
\tau_{n}=\left(D_{\xi}^{2}+{\frac{2 }{3}}v+{\frac{1 }{3}}v_{\xi}
D_{\xi}^{-1}\right)^{n}v_{\xi},  \label{c1}
\end{equation}
are given as follows:
\begin{gather}
\tau_{0}=v_{\xi},  \nonumber \\
\tau_{1}=v_{3\xi}+vv_{\xi},  \nonumber \\
\tau_{2}=v_{5\xi}+{\frac{10 }{3}}v_{\xi} v_{2\xi} +{\frac{5 }{3}}v
v_{3\xi} +{\frac{5 }{6}}v^{2} v_{\xi},  \nonumber \\
\tau_{3}=v_{7\xi}+{\frac{21 }{3}}v_{\xi} v_{4\xi} +{\frac{7 }{3}}v
v_{5\xi} +{\frac{35 }{3}}v_{2\xi} v_{3\xi}  \nonumber \\
\phantom{\tau_{3}=}{}+{\frac{70 }{9}}v v_{\xi} v_{2\xi} +{\frac{35 }{18}}v^{2} v_{3\xi} +{\frac{
35 }{18}} v^{3}_{\xi} +{\frac{35 }{54}}v^{3} v_{\xi}.
\end{gather}
We see, from (\ref{a7}), that differential operators are connected $
D_{\xi}=t^{1/2} D_{x}$. Now, using (\ref{a7}) and (\ref{a8}) directly or the
transformation \cite{olv} ${\cal R}_{\rm cKdV}= \chi_{v} {\cal R}_{\rm KdV}
\chi_{v}^{-1}$, where $\chi_{v}=\frac{\delta u}{\delta v}=1/t$ we transform
the recursion operator of the KdV to the recursion operator of the cKdV
\[
{\cal R}=tD_{x}^2+{\frac{2 }{3}} t u+{\frac{1 }{3}} x+{\frac{1 }{6}} (1+2t
u_{x}) D_{x}^{-1}
\]
and the first four symmetries of cKdV as
\begin{gather}
\sigma_{0}=t^{1/2}u_{x}+{\frac{1 }{2}} t^{-1/2},  \nonumber \\
\sigma_{1}=t^{3/2}u_{t}+t^{1/2}\left(u+{\frac{x }{2}}u_{x}\right)+t^{-1/2}{\frac{x }{4
}} ,  \nonumber \\
\sigma_{2}=t^{5/2}\left(u_{5x}+{\frac{5 }{3}}uu_{3x} +{\frac{10 }{3}}u_{x}
u_{2x} +{\frac{5 }{6}}u^{2} u_{x}\right)  \nonumber \\
\phantom{\sigma_{2}=}{}+t^{3/2}\!\left({\frac{5 }{6}}xu_{3x}+{\frac{5 }{3}}u_{2x}
+{\frac{5 }{6}}xuu_{x}
+{\frac{5 }{12}}u^{2}\right)\!+t^{1/2}\!\left( {\frac{5 }{12}}xu +{\frac{5 }{24}}x^{2}u_{x}\right)\!
+t^{-1/2}{\frac{5x^{2} }{48}},  \nonumber \\
\sigma_{3}=t^{7/2}\left(u_{7x}+{\frac{7 }{3}}u u_{5x}+7u_{x} u_{4x} +{\frac{35
}{18}}u^{2} u_{3x} +{\frac{35 }{3}}u_{2x} u_{3x}
+{\frac{70 }{9}}u u_{x} u_{2x} +{\frac{35 }{18}}u_{x}^{3}\right.
 \nonumber \\ \left.
\phantom{\sigma_{3}=}{}
+{\frac{35 }{54}}
u^{3}u_{x}\right) +t^{5/2}\left({\frac{7 }{6}}xu_{5x} +{\frac{7 }{2}} u_{4x} +{\frac{35
}{18}}xu u_{3x}
+{\frac{35 }{9}}xu_{x} u_{2x} + {\frac{35 }{9}}u u_{2x} +{\frac{35 }{12}}
u_{x}^{2} \right.
\nonumber \\ \left.
\phantom{\sigma_{3}=}{}
+{\frac{35 }{36}}x u^{2}u_{x} + {\frac{35 }{108}}u^{3}\right)+t^{3/2}\left({
\frac{35 }{72}}x^{2}u_{3x}
+{\frac{35 }{18}}xu_{2x}
+{\frac{35 }{72}}x^{2}uu_{x} +{\frac{35 }{24}}
u_{x}\right.
\nonumber \\ \left.
\phantom{\sigma_{3}=}{}
+{\frac{35 }{72}}xu^{2}\right)
+t^{1/2}\left({\frac{35 }{432}}x^{3}u_{x} +{\frac{
35 }{144}}x^{2}u  +{\frac{35 }{144}}\right) +t^{-1/2}{\frac{35 }{864}}x^{3}.
\end{gather}
Here point transformations give the standard (weak) recursion
operator and correct symmetries of the cKdV equation. Therefore
this point transformation shows that adding a~correction term to
the recursion operators is necessary. It means that under point
transformations the main property {\it mapping symmetries to
symmetries} of the recursion operators should be kept invariant.
This is the {\it covariance}. Recursion operators obtained from
(\ref{a1}) do not in general obey this covariance principle.
Another point is that the action of the operator $D^{-1}$ in $(\xi
, \tau)$ system (the KdV case) and in ($t,x$) system (the cKdV
case) should not be the same. In the following sections we
investigate the corrected recursion operators and the behavior of
symmetries and develop a procedure for constructing corrected
symmetries  generated by weak time-dependent recursion operators.

\section{Violation of associativity\\
 and the correct recursion operators}

The symmetries $\sigma \in A$ of the equation (\ref{a0}) obey the evolution
equation (\ref{a2}) and the recursion operator maps symmetries to symmetries,
i.e., ${\cal R} \sigma =\lambda \sigma$.
We assume that a weak recursion operator satisfying (\ref{a1})  takes the
following form ${\cal R}_{w}=R_{1}+a D^{-1}$, where~$R_{1}$ is the local
part of the recursion operator and $a$ is a function of jet coordinates and~$x$ and~$t$.
This is a specific example, a recursion operator may take more
complicated nonlocal terms (Example~8). Now we let ${\cal R} ={\cal R}_{w}+
{\frac{a }{g}} H$, where $H$ is an operator and the function $g$ is chosen so
that ${{a }/{g}}$ is a symmetry. Hence the eigenvalue equation becomes
\begin{equation}
{\cal R}_{w} \sigma+{\frac{a }{g}} H \sigma=\lambda \sigma.
\end{equation}
\noindent Taking the time derivative of the eigenvalue equation and paying
attention to the order of the parenthesis we obtain
\begin{equation}
{\cal R}_{wt} (\sigma)+{\cal R}_{w}\, (K^{\prime}(\sigma))+ (a /g)_{t} H
\sigma +(a / g) (H\sigma)_{t} =K^{\prime}({\cal R}_{w} (\sigma)+(a / g)
H(\sigma)).
\end{equation}
\noindent Since, in general, ${\cal R}_{w}$ contains $D^{-1}$ one should be
careful about the parenthesis. For this reason we rewrite the above equation
as
\begin{equation}
a\,(H(\sigma))_{t}+g [{\rm As}\,({\cal R}_{w},K^{\prime},\sigma)- {\rm As}\,(K^{\prime},
{\cal R}_{w},\sigma)]=0,  \label{r11}
\end{equation}
 where ${\rm As}\, (P,Q,\sigma)=P(Q(\sigma))-(PQ)(\sigma)$ for any operators
$P$, $Q$ and for any $\sigma \in A$. The above equation is the general form
of the additional (correction) term.

For local cases the associators ${\rm As}\,(A,B,\sigma)$ vanish identically. To
correct the symmetries one needs to add only a time-dependent constant, hence
the operator $H$ contains a projection operator $\Pi$ such that $\Pi \sigma=
\underset{x,q,q_{x}, \cdots \rightarrow 0}{\rm Lim} \sigma =$ a time-dependent function.
Hence~(\ref{r11}) reduces to
\begin{equation}
(H\sigma_{0})_{t}+g\left[{\rm As}\left(D^{-1},K_{0}^{\prime},
\sigma_{0}\right)- {\rm
As}\left(K_{0}^{\prime},D^{-1},\sigma_{0}\right)\right]=0,
\label{r12}
\end{equation}
 where $\sigma_{0}$ is the part of the symmetries which depends
only on $x$ and $t$ and $K_{0}^{\prime}=\underset{q,q_{x}, \cdots \rightarrow 0}{\rm Lim}
K^{\prime}$.

\begin{example}[Burgers equation] $u_{t}=K[u]=u_{2x}+uu_{x}$,
$K_{0}^{\prime}=D^2$. For the well known recursion operator ${\cal R}=D+
{\frac{u }{2}}+ {\frac{u_{x} }{2}} D^{-1}$ there is no problem, ${\cal R}=
{\cal R}_{w}$. But if we choose the recursion operator ${\cal R}_{w}=tD +
{\frac{t }{2}} u+{\frac{1 }{2}}x +{\frac{1 }{2}} (1+t u_{x}) D^{-1}$ , there
is a problem in the calculation of symmetries. Here $a={\frac{1 }{2}}
(1+tu_{x})$.  Since $a/g$ is a symmetry, then $g$ must take the value $g=1$.
Let $\sigma_{0}=a_{1}(t)+a_{2}(t) x + a_{3}(t) x^2 + \cdots$ then (\ref{r12}) becomes
\begin{equation}
(H\sigma_{0})_{t}= a_{2}.
\end{equation}
Hence
\begin{equation}
H= D_{t}^{-1}\,\Pi D.  \label{h0}
\end{equation}
\end{example}

\begin{example}[cKdV equation] $u_{t}=K[u]=u_{xxx}+uu_{x}-
{\frac{u }{2t}}$, $K^{\prime}_{0}=D^3-{\frac{1 }{2t}}$. The recursion
operator is ${\cal R}_{w}=tD^2+{\frac{2 }{3}}tu+{\frac{1 }{3}}x + {\frac{1 }{6}}
( 1 +2tu_{x}) D^{-1}$. Here $a={\frac{1 }{6}}(1+2t u_{x})$, and $g=\sqrt{t
}$. Using the same ansatz for $\sigma_{0}$ as in the above example we obtain
\begin{equation}
(H\sigma_{0})_{t}=2 g(t) a_{3}.
\end{equation}
This means that
\begin{equation}
H=D_{t}^{-1}\, \sqrt{t} \Pi\, D^2.  \label{h1}
\end{equation}
The results are compatible with symmetry calculations in the
following sections and also with~\cite{san}. One might generalize the
formula (\ref{r12}) for more complicated evolution  equations with $\underset{q
\rightarrow 0}{\rm Lim}\,K^{\prime} \ne 0 $.
\end{example}

\section{Construction of symmetries}
Firstly we look at the action of $D^{-1}$ on local functions. Let $G \in
A_{1}$.  Then we take $D^{-1} G_{x}=G$ and let $H \in A_{0}$.  Then we take
 $D^{-1} H_{x}=H+h(t)$, where $h$ is a function of~$t$. We start with the
following definition.

\begin{definition} Let ${\cal R}_{w}$ be a
recursion operator of the form
\begin{equation}
{\cal R}_{w}=R_{1}+R_{0},  \label{c3}
\end{equation}
\noindent where $R_{0}={\cal R}_{w}|_{q \rightarrow 0}$. Here and in the
sequel $q \rightarrow 0$ means all the derivatives of $q$ also go to zero
(jet coordinates vanish).
 Similarly, let $\sigma_{n}$ be symmetries of (\ref{a0}), generated
by the~${\cal R}_{w} $, of the form
\begin{equation}
\sigma_{n}= \sigma^{1}_{n}+\sigma^{0}_{n},
\end{equation}
where $\sigma^{0}_{n}=\sigma_{n}|_{q=0}$.
\end{definition}

 At this point we need the following proposition.

\begin{proposition} Let the function $K$
vanish in the limit when the jet coordinates go to zero, i.e.\
$\underset{q
\rightarrow 0}{\rm Lim}\, K=0$. Then the operator $R_{0}=\underset{q \rightarrow 0}{\rm Lim}\,
 {\cal R}_{w}$ satisfies $\sigma^{0}_{n+1}=R_{0}\sigma^{0}_{n}$ and $%
R_{0t}=[K^{\prime}_{0},R_{0}]$ where $K^{\prime}_{0}$ is obtained from (\ref
{a0}) by $K^{\prime}_{0}=\underset{q \rightarrow 0}{\rm Lim}\, K^{\prime}$.
\end{proposition}

We omit the proof because it is straightforward. The difference
between the weak symmetries (the ones obtained from ${\cal R}
_{w}$) and the corrected symmetries comes from $\sigma_{0}$ part
of the symmetries. For this purpose this proposition will play an
important role in the calculation of the missing terms in the
symmetries. When we find the correction term~$h(t)$ for
$\sigma_{0}$ the general corrected  symmetry $\sigma$ takes the
form
\begin{equation}
\sigma= \bar {\sigma}+{\frac{a }{g}} h,  \label{t0}
\end{equation}
where $\bar {\sigma}$ is the one obtained by the weak recursion
operator. The corresponding corrected recursion operator takes the form
\begin{equation}
{\cal R}= {\cal R}_{w}+{\frac{a }{g}}\,H,  \label{r0}
\end{equation}
\noindent and $h=H\,\sigma$. Let us illustrate the procedure of how to
construct the symmetries of an equation from a time-dependent recursion
operator. Firstly we consider the scalar evolution equations of the form $u_{t}=K[u]$.

\begin{example}
 The Burgers equation
\begin{equation}
u_{t}=u_{2x}+uu_{x},  \label{m0}
\end{equation}
\noindent possesses a recursion operator of the form
\begin{equation}
{\cal R}_{w}=tD + {\frac{1 }{2}} t u+{\frac{1 }{2}}x +{\frac{1 }{2}} (1+t
u_{x}) D^{-1},
\end{equation}
where
\begin{equation}
R_ {0}=tD +{\frac{1 }{2}}x +{\frac{1 }{2}}D^{-1}.
\end{equation}
Let
\begin{equation}
\sigma^{0}_{n}= a_{1}+a_{2}x+ a_{2}x^{2}+a_{3}x^{3}+\cdots,
\end{equation}
 where $a_{i}$ are some functions of $t$. From the linearized equation $
\sigma^{0}_{(n)t} =\sigma^{0}_{(n)2x}$
\begin{equation}
a_{1t}=2a_{3}, \qquad a_{2t}=6a_{4}, \qquad
a_{3t}=12a_{5} ,\quad \ldots
\label{a4}
\end{equation}
and by Proposition~1 we obtain
\begin{gather}
\sigma^{0}_{n+1}=\left(tD+{\frac{1 }{2}}x +{\frac{1 }{2}}D^{-1}\right)\sigma^{0}_{n}
= t\left(a_{2}+2a_{3}x +3a_{4}x^{2}+\cdots \right)  \nonumber \\
\phantom{\sigma^{0}_{n+1}=}{}+{\frac{1 }{2}}x(a_{1}+a_{2}x+a_{3}x^{2}+\cdots)
+{\frac{1 }{2}}\left(a_{1}x+{\frac{1 }{2}}a_{2}x^{2}+\cdots +h(t)\right),
\end{gather}
or
\begin{equation}
\sigma^{0}_{n+1}=\left(ta_{2}+{\frac{1 }{2}}h\right)+(12ta_{3}+a_{1})x +\left(3a_{4}t +
{\frac{3 }{4}}a_{2}\right)x^{2}+\cdots .
\end{equation}
Using $\sigma^{0}_{(n+1)t} =\sigma^{0}_{(n+1)2x}$ and equating the
coefficients at power of $x$ to zero we obtain the following system of
equations for $a_{i}$ and $h$
\begin{gather}
\left(ta_{2}+{\frac{1 }{2}}h\right)_{t}= 2\left(3a_{4}t+ {\frac{3 }{4}}a_{2}\right),\nonumber \\
\left(2ta_{3}+a_{1}\right)_{t}= 6 \left(4a_{5}t+ {\frac{2 }{3}}a_{3}\right),\nonumber \\
 \cdots \cdots \cdots\cdots\cdots\cdots\cdots\cdots\cdots\cdots\cdots
\end{gather}
With (\ref{a4}) the first equation gives $h_{t}= a_{2}$ and all
the others are satisfied identically.  Finally we may write $h$ as
\begin{equation}
h=D^{-1}_{t}\left(\Pi D \sigma^{0}_{n}\right),
\end{equation}
\noindent where $\Pi$ is the projection operator defined as $\Pi \,
h(x,t,u,u_{x},\ldots )=h(t,0,0, \ldots) $ for any function $h$. This calculation allows us to write
\begin{equation}
\sigma^{0}_{n+1} =\bar{\sigma}^{0}_{n+1} +{\frac{1 }{2}}D_{t}^{-1}\left(\Pi D
\sigma^{0}_{n}\right),
\end{equation}
where $\bar{\sigma}^{0}_{n+1}$ is the standard part of $
\sigma^{0}_{n+1}$ without the constant of integration $h(t)$. This means
that one should add this constant of integration to $D^{-1}$ in the general
symmetry equation (\ref{t0}),
\begin{equation}
\sigma_{n+1} =\bar{\sigma}_{n+1} +{\frac{1 }{2}}(1+tu_{x})D^{-1}_{t}\left(\Pi D
\sigma^{0}_{n}\right),  \label{a10}
\end{equation}
to allow one to generate the whole hierarchy of symmetries. Here $\bar{\sigma}
_{n+1}$ is the symmetry obtained by standard application of the operator $
D^{-1}$. The corresponding corrected recursion operator (\ref{r0}) for (\ref{m0}) is
\begin{equation}
{\cal R}={\cal R}_{w} +{\frac{1 }{2}}(1+tu_{x})D^{-1}_{t} \Pi D.
\end{equation}
\end{example}

\begin{example} The cylindrical Korteweg-de Vries
equation (cKdV). The cKdV equation,
\begin{equation}
u_{t} =u_{3x}+uu_{x}-{\frac{u }{2t}},  \label{m1}
\end{equation}
possesses a recursion operator of the form
\begin{equation}
{\cal R}_{w}=tD^{2}+{\frac{2 }{3}}tu +{\frac{1 }{3}}x+ {\frac{1 }{6}}
(1+2tu_{x})D^{-1}.  \label{c0}
\end{equation}
The $u$-independent part of recursion operator is
\begin{equation}
R_{0}=tD^{2}+{\frac{1 }{3}}x +{\frac{1 }{6}}D^{-1}.
\end{equation}
Let $\sigma^{0}_{n}=a_{1}+a_{2}x +a_{3}x^{2}+a_{4}x^{3}+\cdots$ with $
\sigma^{0}_{nt}=\sigma^{0}_{n3x}-{\frac{1 }{2t}}\sigma^{0}_{n}$. From these
we have the following relations among the parameters $a_{1t}=6a_{4}-
{\frac{a_{1} }{2t}}$, $a_{2t}=24a_{5}-{\frac{a_{2} }{2t}}$,
$a_{3t}=60a_{6}-{\frac{a_{3} }{2t}}$. Then
\begin{gather}
\sigma^{0}_{n+1}= R_{0}\sigma^{0}_{n}
=\left(2ta_{3}+{\frac{1 }{6}}h\right)+{\frac{1 }{2}}a_{1}x+\left(12ta_{5}+ {\frac{5 }{12}}
a_{2}\right)x^{2}+\cdots.
\end{gather}
Using the linearized equation satisfied by $\sigma^{0}_{n+1}$ we
obtain an equation for $h$ $h_{t}+{\frac{1 }{2t}}h=2a_{3}$ which
gives $h={\frac{1 }{\sqrt{t}}}D^{-1}_{t}\left(\sqrt{t}\Pi
D^{2}\sigma^{0}_{n}\right)$. Hence the symmetry equation
(\ref{t0}) for cKdV equation is
\begin{equation}
\sigma_{n+1}=\bar{\sigma}_{n+1}+{\frac{1 }{6}}(2tu_{x}+1) {\frac{1 }{\sqrt{t}
}}D^{-1}_{t}\left(\sqrt{t} \Pi D^{2}\sigma^{0}_{n}\right).
\end{equation}
The corresponding corrected recursion operator (\ref{r0}) for (\ref{m1}) is
\begin{equation}
{\cal R}={\cal R}_{w} +{\frac{1 }{6}}(1+2tu_{x}){\frac{1 }{\sqrt{t}}}
D^{-1}_{t} \sqrt{t} \Pi D^{2}.
\end{equation}
\end{example}

Now we discuss the symmetries of scalar evolution equation of the
following form
\begin{equation}
u_{t}=F[u]+g(x,t),  \label{z6}
\end{equation}
where $\underset{u \rightarrow 0}{\rm Lim}\, F=0 $ and $g(x,t)$ is an explicit $x$
and $t$ dependent (differentiable) function. We assume that the above
equation admits a recursion operator of the form (\ref{c3}) and any symmetry
of this equation has the form $\sigma_{n}=\sigma^{0}_{n}+\sigma^{1}_{n}+
\sigma^{2}_{n}$, where $\sigma^{0}_{n}=\sigma_{n}|_{u=0}$ and $
\sigma^{1}_{n}=\sum\limits_{i=0} b_{i}u_{i}$ and $\sigma^{2}_{n}$ is the nonlinear
part of the symmetry. Here $b_{i}$ are functions of $x$ and $t$ and $
i=1,2,\ldots$. Now we give the following Proposition.

\begin{proposition} The operator $R_{0}$,
such that $\sigma^{0}_{n+1} =R_{0}\sigma^{0}_{n}$, can be shown to satisfy $
R_{0t}+ {\cal R}_{wt}|_{u \rightarrow 0} =[F^{\prime}_{0} ,R_{0}]$ and the
equation for $\sigma^{0}_{n}$ is
\begin{equation}
\sigma^{0}_{nt}+\sum_{i=0}b_{i}g_{i}=F^{\prime}_{0}\sigma^{0}_{n},
\label{c4}
\end{equation}
\noindent where $F^{\prime}_{0}=F^{\prime}|_{u=0}$.
\end{proposition}

\begin{example} Consider the equation
\cite{cho}
\begin{equation}
u_{t} =u_{3x}+6uu_{x}-{\frac{3u }{t}}-{\frac{5x }{2t^{2}}}  \label{c5}
\end{equation}
which possesses a recursion operator of the form
\begin{equation}
{\cal R}_{w}=t^{6}D^{2}+4t^{6}u +2xt^{5}+ t^{5}(1+2tu_{x})D^{-1}.
\end{equation}
Here $R_{0}=t^{6}D^{2}+2xt^{5}+t^{5} D^{-1}$ satisfies the
relation $R_{0}\sigma^{0}_{n}=\sigma^{0}_{n+1}$. By Proposition 2
the linearized equation for $\sigma^{0}_{n}$ is
\begin{equation}
\sigma^{0}_{nt}-{\frac{5xb_{0} }{2t^{2}}}- {\frac{5b_{1} }{2t^{2}}} =
\sigma^{0}_{n3x}-{\frac{3\sigma^{0}_{n} }{t}}.
\end{equation}
Now taking
\begin{equation}
b_{0}=b^{0}_{0}+b^{1}_{0}x+b^{2}_{0}x^{2}+\cdots,\qquad
b_{1}=b^{0}_{1}+b^{1}_{1}x +b^{2}_{1}x^{2}+\cdots,
\end{equation}
where $b^{i}_{j}=b^{i}_{j}(t)$, we obtain the following equations
\begin{equation*}
a_{0t}=6a_{3}-{\frac{3a_ {0} }{t}}+{\frac{5b^{0}_ {1} }{2t^{2}}}, \qquad
a_{1t}=24a_{4}-{\frac{3a_ {1} }{t}}+{\frac{5b^{0}_ {0} }{2t^{2}}} +{\frac{
5b^{1}_ {1} }{2t^{2}}},\qquad \ldots
\end{equation*}
The next symmetry $\sigma^{0}_{n+1}$ can be generated by the
operator $R_{0}$ and satisfies the following linearized equation according
to Proposition~2
\begin{equation}
\sigma^{0}_{(n+1)t}-{\frac{5xh_{0} }{2t^{2}}}- {\frac{5h_{1} }{2t^{2}}} =
\sigma^{0}_{(n+1)3x}-{\frac{3\sigma^{0}_{n+1} }{t}},  \label{c6}
\end{equation}
 where
\begin{equation}
h_{0}=h^{0}_{0}+h^{1}_{0}x +h^{2}_{0}x^{2}+\cdots,\qquad
h_{1}=h^{0}_{1}+h^{1}_{1}x +h^{2}_{1}x^{2}+\cdots,
\end{equation}
and $h^{i}_{j}=h^{i}_{j}(t)$ . The relation between the $h^{i}_{j}$
and $b^{i}_{j}$ is given by $\sigma_{n+1}={\cal R}_{w}\sigma_{n}$. Now
using~(\ref{c6}) we obtain the equation satisfied by the constant of integration $f$
\begin{equation}
-2t^{7}f_{t}+4t^{7}a_{3}-16t^{6}f+5h^{0}_{0}-10t^{6}\left(h^{2}_{1} -h^{1}_{0}\right)=0,
\end{equation}
where
\begin{equation}
h^{0}_{0}=2t^{6}f+2t^{6}\left(h^{2}_{1} -h^{1}_{0}\right)+t^{6}\sum_{i=2} (-1)^{i}\Pi
D^{i-2}{\frac{\partial \sigma_{n} }{\partial u_{i}}}.
\end{equation}
Hence the constant of integration $f$ becomes
\begin{equation}
f={\frac{1 }{t^{3}}} \left[D^{-1}_{t}\left(\tau^{3} \Pi D^{2}\sigma_{n}\right)+{\frac{5
}{2}} D^{-1}_{t}\left(\tau\sum_{i=2}\Pi D^{i-2}{\frac{\partial \sigma_{n} }{
\partial u_{i}}}\right) \right].
\end{equation}
Therefore the corrected symmetry equation of (\ref{c5}) is of the
form
\begin{gather}
\sigma_{n+1}=\bar{\sigma}_{n+1}+t^{2}(1+2tu_{x}) \nonumber\\
\phantom{\sigma_{n+1}=}{}\times\left[ D^{-1}_{t}\left(\tau^{3}
\Pi D^{2}\sigma_{n}\right)+{\frac{5 }{2}} D^{-1}_{t}\left(\tau\sum_{i=2} (-1)^i\,\Pi
D^{i-2} {\frac{\partial \sigma_{n} }{\partial u_{i}}}\right) \right].
\end{gather}
The first four symmetries of (\ref{c5}) are:
\begin{gather}
\sigma_{0}=t^{2}+2t^{3}u_{x},  \nonumber \\
\sigma_{1}=2t^{9}(u_{3x}+6uu_{x})+6t^{8}(xu_{x}+u) +3t^{7}x,  \nonumber \\
\sigma_{2}=2t^{15}\left(u_{5x}+10uu_{3x} +20u_{x} u_{2x} +30u^{2}u_{x}\right)
\nonumber \\
\phantom{\sigma_{2}=}{}+10t^{14}\left(xu_{3x} +2u_{2x}+6xuu_{x}+3u^{2}\right)
+15t^{13}\left( 2xu +x^{2}u_{x}\right)+{
\frac{15 }{2}}t^{12}x^{2},  \nonumber \\
\sigma_{3}=2t^{21}\left(u_{7x}+14u u_{5x}+42u_{x}u_{4x}
+70u_{2x}u_{3x}+70u^{2}u_{3x}+280uu_{x} u_{2x} \right. \nonumber \\
\left.\phantom{\sigma_{3}=}{}+140u^{3} u_{x} +70u_{x}^{3}\right)+
14t^{20}(xu_{5x}+3u_{4x}+10xuu_{3x} +20xu_{x}u_{2x}+20uu_{2x}
\nonumber \\
\left.\phantom{\sigma_{3}=}{}+15u_{x}^{2}+30xu^{2}u_{x}
+10u^{3}\right)
+35t^{19}
\left(x^{2} u_{3x} +4x u_{2x}+6x^{2}uu_{x}+3u_{x}+6xu^{2}\right)  \nonumber\\
\phantom{\sigma_{3}=}{}+{\frac{35 }{2}}t^{18}\left(2x^{3}u_{x}+6x^{2}u+1\right)
+{\frac{35 }{2}}t^{17}x^{3}.
\end{gather}
\end{example}

 In the previous examples the nonlocal parts of the recursion
operators of evolution equations are of the form $aD^{-1}$. But there exist
some integrable evolution equations admitting recursion operators in which
nonlocal parts are of the form $aD^{-1}b$ where $a$ and~$b$ are in general
functions of both jet coordinates and explicitly $x$ and $t$. In this case,
we may define the nonlocal part of $R_{0}={\cal R}_{w}|_{u \rightarrow 0}=a
h(t)$ when ${b \rightarrow 0}$ as ${u \rightarrow 0}$.

\begin{example} Consider the following extended potential KdV
(pKdV) equation \cite{kar3}
\begin{equation}
u_{t} =u_{3x}+u_{x}^{2}+c_{0}x+c_{1},  \label{z1}
\end{equation}
where $c_{0}$ and $c_{1}$ are arbitrary constants. The recursion
operator for (\ref{z1}) is given by
\begin{equation}
{\cal R}_{w}= D^{2}+{\frac{4 }{3}}u_{x} -{\frac{4 }{3}}c_{0}t -{\frac{2 }{3}}
D^{-1}u_{2x}.
\end{equation}
As we mentioned above, the form of $R_{0}$ will be
\begin{equation}
R_{0}= D^{2}-{\frac{4 }{3}}c_{0}t-{\frac{2 }{3}}h(t).
\end{equation}
 We observe that the $u$-independent part ($\sigma^{0}_{n}$) of one set
of symmetries are functions of $t$. Therefore from Proposition~2 the
linearized equation for $\sigma^{0}_{n}$ is
\begin{equation}
\sigma^{0}_{nt}+b^{n}_{1} c_{0}= 0,  \label{p0}
\end{equation}
where $b^{n}_{1}$ depends only on $t$. The next symmetry
$\sigma^{0}_{n+1}$ may be generated by $R_{0}$
\begin{equation}
\sigma^{0}_{n+1}=-{\frac{2 }{3}}(2c_{0}t+h)\sigma^{0}_{n}  \label{z2}
\end{equation}
and satisfies
\begin{equation}
\sigma^{0}_{(n+1)t}+ b^{n+1}_{1} c_{0}= 0,  \label{z3}
\end{equation}
where $b^{n+1}_{1}$ depends only on $t$. Furthermore the relation
between $b^{n}_{1}$ and $b^{n+1}_{1}$ is given by $\sigma_{n+1}={\cal R}
_{w}\sigma_{n}$ as $b^{n+1}_{1}={\frac{2 }{3}}\sigma^{0}_{n}-{\frac{4 }{3}}
c_{0}tb^{n}_{1}$. We can find, together with (\ref{p0}), (\ref{z2})
and~(\ref{z3}), an equation for constant of integration $h$
\begin{equation}
h_{t}\sigma^{0}_{n}+ c_{0}D^{-1}_{t} \sigma^{0}_{n}=0  \label{z4}
\end{equation}
in terms of which (\ref{z2}) becomes
\begin{equation}
\sigma^{0}_{n+1}=-{\frac{2 }{3}}\left(2c_{0}t-c_{0}D^{-1}_{t}\right)\sigma^{0}_{n}.
\end{equation}
This leads to the general symmetry equation
\begin{equation}
\sigma_{n+1}=\bar{\sigma}_{n+1}+{\frac{2 }{3}}c_{0} D^{-1}_{t}\Pi
\sigma^{0}_{n}.
\end{equation}
The first four symmetries (\ref{z1}) are:
\begin{gather}
\sigma_{0}=1,\nonumber\\
 \sigma_{1}={\frac{2 }{3}}(u_{x}-c_{0}t),  \nonumber \\
\sigma_{2}={\frac{2 }{3}}\left(u_{3x}+u_{x}^{2}-2c_{0}tu_{x}+c_{0}^{2}t^{2}\right),
\nonumber \\
\sigma_{3}={\frac{2 }{27}}\left(9u_{5x}+30u_{x}u_{3x}-30c_{0}tu_{3x}
+15u_{2x}^{2} \right. \nonumber \\
\left.\phantom{\sigma_{3}=}{}+10u_{x}^{3}-30c_{0}tu_{x}^{2}+30c_{0}^{2}t^{2}u_{x}-10c_{0}^{3}t^{3} \right).
\end{gather}
\end{example}

We finally remark that the corrected recursion operator for (\ref
{z1}), taking into account the symmetry structure of (\ref{z1}), may be
written as
\begin{equation}
{\cal R}= D^{2}+{\frac{4 }{3}}u_{x} -{\frac{4 }{3}}c_{0}t -{\frac{2 }{3}}
D^{-1}u_{2x}+{\frac{2 }{3}}c_{0}D^{-1}_{t}\Pi.
\end{equation}

In the following section we will consider the time-dependent symmetries of a
system of evolution equations.

\section{System of evolution equations}

Following the procedure introduced in Section~4, we now discuss the
time-dependent symmetries for a system of evolution equations. We present
several examples.

\begin{example} Consider the following
nonautonomous system of equations \cite{kar2}
\begin{gather}
 u_{t}=u_{3x},  \nonumber \\
 v_{t}= v_{3x}+{\frac{c_{0} }{\sqrt {t}}}uu_{x},  \label{d7}
\end{gather}
 with recursion operator
\begin{equation}
{\cal R}_{w}= \left(
\begin{array}{cc}
\displaystyle tD^{2}+{\frac{x }{3}}+{\frac{1 }{6}}D^{-1} & 0 \vspace{2mm}\\
\displaystyle {\frac{2c_{0} \sqrt {t} }{3}}u+{\frac{c_{0} \sqrt {t} }{3}}u_{x}D^{-1} &
\displaystyle tD^{2}+{\frac{x }{3}}+{\frac{1 }{6}}D^{-1}
\end{array}
\right),  \label{c8}
\end{equation}
where $c_{0}$ is an arbitrary constant.  In a similar way as for
scalar evolution equations we may take the form of symmetries $
\sigma^{0}_{n}$ and $\psi^{0}_{n}$ as
\begin{gather}
\sigma^{0}_{n}= a_{1}+a_{2}x+ a_{2}x^{2}+a_{3}x^{3}+\cdots,  \nonumber \\
\psi^{0}_{n}= b_{1}+b_{2}x+ b_{2}x^{2}+b_{3}x^{3}+\cdots,  \label{d0}
\end{gather}
where $a_{i}$ and $b_{i}$ are some functions of $t$. The
linearized equations, by Proposition~1, for those symmetries, viz.
\begin{equation}
\left( \begin{array}{c}
\sigma^{0}_{n} \vspace{1mm}\\
\psi^{0}_{n}
\end{array}
\right)_{t} =\left(
\begin{array}{c}
\sigma^{0}_{n} \vspace{1mm}\\
\psi^{0}_{n}
\end{array}
\right)_{3x},   \label{c9}
\end{equation}
with a simple comparison of each power of $x$, lead to
\begin{gather}
a_{1t}=6a_{4}, \qquad a_{2t}=24a_{5}, \qquad a_{3t}=60a_{6} , \qquad \ldots ,
\nonumber \\
b_{1t}=6b_{4}, \qquad b_{2t}=24b_{5}, \qquad b_{3t}=60b_{6} ,
\qquad \ldots . \label{c10}
\end{gather}
The next symmetry, generated by $R_{0}$, is
\begin{equation}
\left(
\begin{array}{c}
\sigma^{0}_{n+1} \vspace{1mm}\\
\psi^{0}_{n+1}
\end{array}
\right) = R_{0} \left(
\begin{array}{c}
\sigma^{0}_{n} \vspace{1mm}\\
\psi^{0}_{n}
\end{array}
\right),  \label{c11}
\end{equation}
where
\begin{equation}
R_{0}= \left(
\begin{array}{cc}
\displaystyle tD^{2}+{\frac{x }{3}}+{\frac{1 }{6}}D^{-1} & 0 \vspace{2mm}\\
0 & \displaystyle tD^{2}+{\frac{x }{3}}+{\frac{1 }{6}}D^{-1}
\end{array}
\right)  \label{e0}
\end{equation}
is the $(u,v)$ independent part of the recursion operator of ${\cal R}_{w}$
satisfies the Proposition~1. Hence
\begin{gather}
\sigma^{0}_{n+1}= \left(2a_{3}t+{\frac{1 }{6}}h\right)
+\left({\frac{1 }{2}}a_{1}+6a_{4}t\right)x
+\left({\frac{5 }{12}}a_{2}+12a_{5}t\right)x^{2}+\cdots ,  \nonumber \\
\psi^{0}_{n+1}= \left(2b_{3}t+{\frac{1 }{6}}g\right)+\left({\frac{1 }{2}}b_{1}+6b_{4}t\right)x +
\left({\frac{5 }{12}}b_{2}+12b_{5}t\right)x^{2}+\cdots ,
\end{gather}
\noindent where $h(t)$ and $g(t)$ are the constants of integrations. Using
linearized equations for $\sigma^{0}_{n+1}$ and $\psi^{0}_{n+1}$
\begin{equation}
\left(
\begin{array}{c}
\sigma^{0}_{n+1}\vspace{1mm}\\
\psi^{0}_{n+1}
\end{array}
\right)_{t} =\left(
\begin{array}{c}
\sigma^{0}_{n+1} \vspace{1mm}\\
\psi^{0}_{n+1}
\end{array}
\right)_{3x}
\end{equation}
together with (\ref{c10}) we find the values of the constants of
integrations $h$ and $g$ as $h(t)=2a_{3}$ and $g(t)=2b_{3}$. Finally we may
write
\begin{equation}
h(t)=D^{-1}_{t}\left(\Pi D^{2} \sigma^{0}_{n}\right) ,\qquad
g(t)=D^{-1}_{t}\left(\Pi D^{2} \psi^{0}_{n}\right),
\end{equation}
where the projection $\Pi$ is defined as $\Pi
h(x,t,u,u_{x},\ldots,v,v_{x}, \ldots)=h(t,0,0,\ldots)$ for any function $h$. The
general symmetry equations (\ref{t0}) for (\ref{d7}) are of the form
\begin{gather}
\sigma_{n+1}= \bar \sigma_{n+1}+{\frac{1 }{6}} D^{-1}_{t}\left(\Pi D^{2}
\sigma^{0}_{n}\right),  \nonumber \\
\psi_{n+1}= \bar \psi_{n+1}+{\frac{2 }{3}} \sqrt {t}c_{0}u_{x}
D^{-1}_{t}\left(\Pi D^{2} \sigma^{0}_{n}\right)+{\frac{1 }{6}} D^{-1}_{t}\left(\Pi D^{2}
\psi^{0}_{n}\right),
\end{gather}
where $\bar \sigma_{n+1}$ and $\bar \psi_{n+1}$ are the symmetries
obtained by standard application of the operator~$D^{-1}$. Let
\begin{equation}
\tau_{n}= \left(
\begin{array}{c}
\sigma_{n} \\
\psi_{n}
\end{array}
\right)
\end{equation}
be the symmetries of (\ref{d7}). Then the first four symmetries of (\ref{d7}) are:
\begin{gather}
\sigma_{0}=1,  \nonumber \\
\psi_{0}=1+2c_{0}t^{1/2}u_{x} ,  \nonumber \\
\sigma_{1}={\frac{1 }{2}}x ,  \nonumber \\
\psi_{1}= 2 t^{3/2}u_{3x}+t^{1/2}c_{0}(xu_{x}+u)+{\frac{1 }{2}}x,
\nonumber \\
\sigma_{2}={\frac{5 }{24}}x^{2},  \nonumber \\
\psi_{2}= 2c_{0}t^{5/2}u_{5x}+c_{0}t^{3/2}\left({\frac{5 }{3}}xu_{3x}+ {\frac{
10 }{3}}u_{2x}\right)+{\frac{5 }{6}}c_{0}t^{1/2}\left(2x^{2}u_{x}+xu\right)+{\frac{5 }{24}}
x^{2},  \nonumber \\
\sigma_{3}= {\frac{35 }{72}}t+ {\frac{35 }{432}}x^{3},  \nonumber \\
\psi_{3}= 2c_{0}t^{7/2}u_{7x}+c_{0}t^{5/2}\left({\frac{7 }{3}}xu_{5x}+
7u_{4x}\right)+35c_{0}t^{3/2}\left({\frac{1 }{36}}x^{2}u_{3x}+{\frac{1 }{9}}xu_{2x}
+{\frac{1 }{12}}u_{x}\right) \nonumber \\
\phantom{\psi_{3}=}{}+35c_{0}t^{1/2}\left({\frac{1 }{216}}x^{3}u_{x}+{\frac{1
}{72}}u\right)+{\frac{35 }{72}} t+{\frac{35 }{432}}x^{3}.
\end{gather}
\end{example}

\begin{example} The system \cite{kar2}
\begin{gather}
u_{t}=u_{3x}+ {\frac{2c_{0} }{\sqrt {t}}}uu_{x},  \nonumber \\
v_{t}= v_{3x}+ {\frac{c_{0} }{\sqrt {t}}}(uv)_{x},  \label{e3}
\end{gather}
is the nonautonomous Jordan Korteweg-de Vries (JKdV), where $c_{0}$
is an arbitrary constant. The recursion operator ${\cal R}_{w}$ for this
system is
\begin{equation}
{\cal R}_{w}= \left(
\begin{array}{cc}
{\cal R}^{0}_{0} & {\cal R}^{0}_{1} \vspace{1mm}\\
{\cal R}^{1}_{0} & {\cal R}^{1}_{1}
\end{array}
\right)  \label{a03}
\end{equation}
with
\begin{gather}
{\cal R}^{0}_{0}  =  tD^{2}+{\frac{1 }{3}}x+{\frac{4c_{0} }{3}}\sqrt{t}u +
{\frac{1 }{6}}\left(4c_{0}\sqrt{t}u_{x}+1\right)D^{-1},  \nonumber \\
{\cal R}^{0}_{1}  =  0,  \nonumber \\
{\cal R}^{1}_{0}  =  {\frac{2c_{0} }{3}}\sqrt{t}v + {\frac{c_{0} }{3}}
\sqrt{t}v_{x}D^{-1} -{\frac{c_{0}^{2} }{9}} uD^{-1}vD^{-1},  \nonumber \\
{\cal R}^{1}_{1}  =  tD^{2}+{\frac{1 }{3}}x+{\frac{2c_{0} }{3}}\sqrt{t}u+ {
\frac{1 }{6}}(2 c_{0}\, \sqrt{t}u_{x}+1)D^{-1}+{\frac{c_{0}^{2} }{9}}
uD^{-1}uD^{-1}.
\end{gather}
Again we take the form of symmetries $\sigma^{0}_{n}$
and $\psi^{0}_{n}$ as in (\ref{d0}) with the $(u,v)$ independent recursion
operator $R_{0}$ which is the same as in (\ref{e0}). Performing the same
steps as in the previous case, we obtain the general symmetry equations (\ref{t0})
for (\ref{e3}) to be
\begin{gather}
\sigma_{n+1}= \bar \sigma_{n+1}+{\frac{1 }{6}}\left(4c_{0} \sqrt {t}\, u_{x}+1\right)
D^{-1}_{t}\left(\Pi D^{2} \sigma^{0}_{n}\right),  \nonumber \\
\psi_{n+1}= \bar \psi_{n+1}+{\frac{1 }{3}} \sqrt {t}c_{0}v_{x}
D^{-1}_{t}\left(\Pi D^{2} \sigma^{0}_{n}\right)-{\frac{1 }{9}} c_{0}^{2}uD^{-1}v
D^{-1}_{t}\left(\Pi D^{2} \sigma^{0}_{n}\right)  \nonumber \\
\phantom{\psi_{n+1}=}{}+\left({\frac{1 }{6}}\left(2c_{0}\sqrt {t} u_{x}+1\right)
+{\frac{1 }{9}}c^{2}_{0}uD^{-1}u\right)
D^{-1}_{t}\left(\Pi D^{2} \psi^{0}_{n}\right).
\end{gather}
Now we list only the first two symmetries because higher order
ones are too long to write down here.
\begin{gather}
\sigma_{0}={\frac{1 }{6}}\left(4c_{0}t^{1/2}u_{x}+1\right),  \nonumber \\
\psi_{0}={\frac{1 }{18}}\left[6c_{0}t^{1/2}\left(u_{x}+v_{x}\right)+2c_{0}^{2}uw+3\right],
\nonumber \\
\sigma_{1}={\frac{1 }{12}}\left[8c_{0}t^{3/2}u_{3x}+4c_{0}t^{1/2}(xu_{x}+u)
+16c_{0}^{2}tuu_{x}\right],  \nonumber \\
\psi_{1}={\frac{1 }{324}}\bigg[108c_{0}t^{3/2}(u_{3x}+v_{3x})+c_{0}t^{1/2}\left[
12c_{0}^{2}hu_{x}+54xu_{x}+54xv_{x}-12c_{0}^{2}fu \right. \nonumber \\
\left.\phantom{\psi_{1}=}{}+12c_{0}^{2}un +24c_{0}^{2}u^{2}w+54u+54v\right]
+3c_{0}t\left[c_{0}wu_{2x}+2c_{0}u_{x}w_{x}+3c_{0}uu_{x}\right.  \nonumber \\
\left.\phantom{\psi_{1}=}{}+5c_{0}vu_{x} +4c_{0}ug +6c_{0}^{2}h +6c_{0}^{2}pu-
6c_{0}^{2}ru+12c_{0}^{2}xuw+27x\right]\bigg],
\end{gather}
 where $p_{x}=xu$, $n_{x}=u^{2}$,
$r_{x}=xv$, $w_{x}=u-v$, $g_{x}=uh$ and $h_{x}=uw$.

More generally the multicomponent nonautonomous JKdV
system \cite{kar2} is
\begin{equation}
q^{i}_{t}=q^{i}_{3x}+ {\frac{1 }{\sqrt{t}}}s^{i}_{j k}q^{j}q^{k}_{x} ,
\qquad i,j,k=1,2,\ldots,N,  \label{e1}
\end{equation}
where $s^{i}_{j k}$ are constants, symmetric in the lower indices
and satisfy the Jordan identities
\begin{equation}
s^{k}_{p r}F^{i}_{\,\,ljk} +s^{k}_{j r}F^{i}_{\,\,lpk} +s^{k}_{j
p}F^{i}_{\,\,lrk}=0,
\end{equation}
with $F^{i}_{\,\,plj}=s^{i}_{j k} s^{k}_{l p} -s^{i}_{l k} s^{k}_{j p}$.
This system possesses a recursion operator
\begin{gather}
{\cal R}^{i}_{wj}= t \delta^{i}_{j}\,D^{2} +{\frac{2 }{3}}\,\sqrt{t} \,
s^{i}_{jk}\,q^{k}+ {\frac{1 }{3}}\delta^{i}_{j}x +\left({\frac{1 }{3}}\,\sqrt{t}
\, s^{i}_{jk}\,q^{k}_{x}+{\frac{1 }{6}} \delta^{i}_{j}\right) D^{-1}  \nonumber\\
\phantom{{\cal R}^{i}_{wj}=}{} + {\frac{1 }{9}} F^{i}_{\,\,lkj}\,q^{l}\,D^{-1}\,q^{k}\,D^{-1}.
\label{e2}
\end{gather}
The time-dependent symmetries can be computed as
\begin{equation}
\tau^{i}_{n+1}=\bar \tau^{i}_{n+1}+\Lambda^{i}_{j} D^{-1}_{t} \Pi
D^{2}\tau^{j}_{n},
\end{equation}
where
\begin{equation}
\Lambda^{i}_{j}={\frac{1 }{3}}\, \sqrt{t}\, s^{i}_{jk}\,q^{k}_{x}+ {\frac{1
}{6}}\, \delta^{i}_{j}+{\frac{1 }{9}}\,F^{i}_{lkj}\,q^{l} D^{-1}\,q^{k}.
\end{equation}
The corrected recursion operator (\ref{r0}) for (\ref{e1}) is
given by
\begin{equation}
{\cal R}={\cal R}_{w}+ \Lambda D^{-1}_{t} \Pi D^{2}.
\end{equation}
\end{example}

\begin{example}
Consider the following system of
equations
\begin{gather}
 u_{t}=v_{3x}+t^{-2/3}(vu_{x}+vv_{x}),  \nonumber \\
 v_{t}= t^{-2/3}vv_{x},  \label{c12}
\end{gather}
with recursion operator
\begin{equation}
{\cal R}_{w}= \left(
\begin{array}{cc}
3t^{1/3}v+x & 3tD^{2}+3t^{1/3}v+D^{-1} \\
0 & 3t^{1/3}v+x
\end{array}
 \right).  \label{c13}
\end{equation}
We take the form of $\sigma^{0}_{n}$ and $\psi^{0}_{n}$ as in (\ref
{d0}) and from the linearized equations for $\sigma^{0}_{n}$ and~$\psi^{0}_{n}$
\begin{equation}
\left(
\begin{array}{c}
\sigma^{0}_{n} \vspace{1mm}\\
\psi^{0}_{n}
\end{array}
\right)_{t} =\left(
\begin{array}{c}
\psi^{0}_{n} \\
0
\end{array}
\right)_{3x},  \label{c14}
\end{equation}
we obtain
\begin{equation}
a_{1}=6b_{4}t,\qquad a_{2}=24b_{5}t,\qquad a_{3}=60b_{6}t,\qquad
a_{4}=120b_{7}t, \qquad \ldots,  \label{l0}
\end{equation}
where $b_{i}$ are arbitrary constants. The next symmetries, generated by $
R_{0}$, are given in (\ref{c11}) where, in this case, $R_{0}$ has the form
\begin{equation}
R_{0}= \left(
\begin{array}{cc}
x & 3tD^{2}+D^{-1} \\
0 & x
\end{array}
 \right).
\end{equation}
They are
\begin{gather}
\sigma^{0}_{n+1}= (2b_{3}t+h)+(a_{1}+b_{1}+18b_{4}t)x +\left(a_{2}+{\frac{1 }{2}
}b_{2}+36b_{5}t\right)x^{2}+\cdots,  \nonumber \\
\psi^{0}_{n+1}= b_{1}x+b_{2}x^{2} +b_{3}x^{3}+\cdots,
\end{gather}
where $h$ is the constant of integration. Now using linearized
equations for $\sigma^{0}_{n+1}$ and $\psi^{0}_{n+1}$ together with (\ref{l0})
we obtain the value of $h=0$. We point out that ${\cal R}_{w}={\cal R}$.
The first three classical symmetries of (\ref{c12}) are:
\begin{equation}
\tau_{0}= \left(
\begin{array}{c}
1 \\
0
\end{array}
\right) , \qquad  \tau_{1}= \left(
\begin{array}{c}
3t^{1/3}v+x \\
0
\end{array}
\right) , \qquad  \tau_{2}= \left(
\begin{array}{c}
(3t^{1/3}v+x)^{2} \\
0
\end{array}
\right) .
\end{equation}
\end{example}

\begin{example} Consider the following system of
equations
\begin{gather}
 u_{t}=u_{3x}+e^{-t}vv_{x},  \nonumber \\
 v_{t}= vv_{x},  \label{c17}
\end{gather}
with recursion operator
\begin{equation}
{\cal R}_{w}= \left(
\begin{array}{cc}
D^{2}+D^{-1} & e^{-t}v \\
0 & v
\end{array}
\right).  \label{c15}
\end{equation}
We take the form of $\sigma^{0}_{n}$ and $\psi^{0}_{n}$ as in
(\ref{d0}) and from the linearized equations (\ref{c14}) for $\sigma^{0}_{n}$ and
$\psi^{0}_{n}$ we obtain
\begin{equation}
a_{1t}=6a_{4},\qquad a_{2t}=24a_{5},\qquad a_{3t}=60a_{6},\qquad
a_{4t}=120a_{7}, \qquad \ldots,  \label{l2}
\end{equation}
where $b_{i}$ are arbitrary constants. The next symmetry, generated by $R_{0}$,
is given in (\ref{c11}), where, in this case, $R_{0}$ has the form
\begin{equation}
R_{0}= \left(
\begin{array}{cc}
D^{2}+D^{-1} & 0 \\
0 & 0
\end{array}
 \right).
\end{equation}
They are
\begin{gather}
\sigma^{0}_{n+1}= (2a_{3}t+h)+(a_{1}+6a_{4}t)x +\left({\frac{1 }{2}}
a_{2}+12a_{5}\right)x^{2}+\cdots ,  \nonumber \\
\psi^{0}_{n+1}= 0,
\end{gather}
where $h$ is the constant of integration. Now using linearized equations for
$\sigma^{0}_{n+1}$ and $\psi^{0}_{n+1}$ together with (\ref{l2}) we obtain
the value of $h=D^{-1}_{t}\left(\Pi D^{2}\sigma^{0}_{n}\right)$.

Hence the symmetry equations of (\ref{c13}) take the following form
\begin{gather}
\sigma_{n+1}= \bar \sigma_{n+1}+ D^{-1}_{t}\left(\Pi D^{2} \sigma^{0}_{n}\right),
\nonumber \\
\psi_{n+1}= \bar \psi_{n+1}.
\end{gather}

The first four symmetries of (\ref{c17}) are:
\begin{equation}
\tau_{0}= \left(
\begin{array}{c}
1 \\
0
\end{array}
\right) , \quad \tau_{1}= \left(
\begin{array}{c}
x \\
0
\end{array}
\right) , \quad  \tau_{2}= \left(
\begin{array}{c}
x^{2} \\
0
\end{array}
\right) , \quad \tau_{3}= \left(
\begin{array}{c}
{\frac{1 }{6}}x^{3}+t+1 \\
0
\end{array}
\right) .
\end{equation}
\end{example}

\begin{example}
Consider the following system of equations
\begin{gather}
u_{t}=u_{3x}+c_{0}v_{3x}-c_{0}(u_{x}v +c_{0}vv_{x}),  \nonumber \\
v_{t}= u_{x}v+c_{0}vv_{x}  \label{c18}
\end{gather}
with recursion operator
\begin{equation}
{\cal R}_{w}= \left(
\begin{array}{cc}
3tD^{2}+x+2 D^{-1}-3c_{0}t v & c_{0}\left(3tD^{2}+x+2 D^{-1}-3c_{0}tv\right) \\
3t v & 3c_{0}tv+x
\end{array}
\right),  \label{c19}
\end{equation}
where $c_{0}$ is an arbitrary constant. We take the form of
$\sigma^{0}_{n}$ and $\psi^{0}_{n}$ as in (\ref{d0}) and from the linearized
equations
\begin{equation}
\left(
\begin{array}{c}
\sigma^{0}_{n} \vspace{1mm}\\
\psi^{0}_{n}
\end{array}
\right)_{t} =\left(
\begin{array}{cc}
1 & c_{0} \\
0 & 0
\end{array}
 \right) \left(
\begin{array}{c}
\sigma^{0}_{n} \vspace{1mm}\\
\psi^{0}_{n}
\end{array}
\right)_{3x}
\end{equation}
we obtain
\begin{equation}
a_{1t}=6a_{4}+6c_{0}b_{4} , \qquad a_{2t}=24a_{5}+24c_{0}b_{5},
\qquad a_{3t}=60a_{5}+60c_{0}b_{6} ,\qquad \ldots,  \label{l3}
\end{equation}
where, in this case, $b_{i}$ are arbitrary constants. The next
symmetry, generated by $R_{0}$, is
\begin{equation}
\left(
\begin{array}{c}
\sigma^{0}_{n+1} \vspace{1mm}\\
\psi^{0}_{n+1}
\end{array}
\right) = R_{0} \left(
\begin{array}{c}
\sigma^{0}_{n} \vspace{1mm}\\
\psi^{0}_{n}
\end{array}
\right),
\end{equation}
where
\begin{equation}
R_{0}= \left(
\begin{array}{cc}
3tD^{2}+x+2 D^{-1} & c_{0} \left(3tD^{2}+x+2 D^{-1}\right) \\
0 & x
\end{array}\right).
\end{equation}
is the $(u,v)$ independent part of the recursion  operator of ${\cal R}_{w}$.
Using the linearized equations for $\sigma^{0}_{n+1}$  and $\psi^{0}_{n+1}$
together with (\ref{l3}) we may determine the value of constant of
integration as
\begin{equation}
h=D^{-1}_{t}\left(\Pi D^{2} \sigma^{0}_{n}\right) + c_{0}t\left(\Pi D^{2} \psi^{0}_{n}\right).
\end{equation}
Therefore the symmetry equations of (\ref{c18}) are:
\begin{gather}
\sigma_{n+1}= \bar \sigma_{n+1}+2(c_{0}+1) \left(D^{-1}_{t}\left(\Pi D^{2}
\sigma^{0}_{n}\right)+ c_{0}t\left(\Pi D^{2} \psi^{0}_{n}\right)\right),  \nonumber \\
\psi_{n+1}= \bar \psi_{n+1}.
\end{gather}
The first four symmetries of (\ref{c18}) are:
\begin{gather}
\sigma_{0}=2(c_{0}+1),\qquad \psi_{0}=0 ,  \nonumber \\
\sigma_{1}=-6c_{0}(c_{0}+1)tv +2(3x(c_{0}+1)+c_{0}) ,\qquad
\psi_{1}= 6(c_{0}+1)tv , \nonumber \\
\sigma_{2}=-6c_{0}tv(4x(c_{0}+1)+c_{0})+6x(2x(c_{0}+1)+c_{0})),  \nonumber\\
\psi_{2}= 6vt(4x(c_{0}+1)+c_{0}) ,  \nonumber \\
\sigma_{3}= -12c_{0}txv(5x(c_{0}+1)+2c_{0})+4(c_{0}+1)\left(30t+5x^{3}\right)
+12c_{0}x^{2}+120t,  \nonumber \\
\psi_{3}= 12xtv(5x(c_{0}+1)+2c_{0}).
\end{gather}
\end{example}

\section{Conclusion}
It is well known that one of the effective ways to find symmetries
is to use the recursion ope\-ra\-tor. If the recursion operator or
the corresponding evolution equation is time-dependent one is
faced with some difficulties in finding the correct symmetries.
Here, in this work, we approach this problem in two ways. Firstly
we observed that by the standard application of $D^{-1}$ on
functions having explicit $t$ and $x$ dependencies the rule of
associativity is lost in the application of consecutive two
operators on such a function space. Due to this fact the standard
equation for the recursion operators is no longer valid. We
modified this equation by including the associators. We have given
some applications of our formula~(\ref{r12}) for the modified term
of the recursion operators. The second way to calculate the
missing parts of the symmetries is to use directly the symmetry
equation and the correct application of~$D^{-1}$ on functions
having explicit $x$ and $t$ dependencies. We have given a general
treatment and several examples.

\subsection*{Acknowledgements}

This work is partially supported by the Scientific and Technical Research
Council of Turkey (TUBITAK) and Turkish Academy of Sciences (TUBA).

\label{gurses-lastpage}
\end{document}